\documentclass[preprintnumbers,amsmath,amssymbm,prd]{revtex4}
\usepackage{epsfig}
\usepackage{graphicx}
\usepackage{amssymb}

\begin{document}
\title{Lower bound on the proper lengths of stationary bound-state charged massive scalar clouds}
\author{Shahar Hod}
\affiliation{The Ruppin Academic Center, Emeq Hefer 40250, Israel}
\affiliation{ } \affiliation{The Jerusalem Multidisciplinary Institute, Jerusalem 91010, Israel}
\date{\today}

\begin{abstract}
It has recently been revealed that charged scalar clouds, 
spatially regular matter configurations which are made of linearized charged massive scalar fields, 
can be supported by spinning and charged Kerr-Newman black holes. 
Using analytical techniques, we establish a no-short hair theorem for these stationary bound-state 
field configurations. 
In particular, we prove that the effective proper lengths of the supported charged massive scalar clouds 
are bounded from below by the remarkably compact dimensionless relation 
$\ell/M>\ln(3+\sqrt{8})$, where $M$ is the mass of the central supporting black hole. 
Intriguingly, this lower bound is universal in the sense that it is valid for all Kerr-Newman black-hole spacetimes 
[that is, in the entire regime $\{a/M\in(0,1],Q/M\in[0,1)\}$ of the dimensionless spin and charge parameters that characterize 
the central supporting black holes] and for all values of the physical parameters 
(electric charge $q$, proper mass $\mu$, and angular harmonic indexes $\{\l,m\}$) that characterize 
the supported stationary bound-state scalar fields.
\end{abstract}
\bigskip
\maketitle

\section {Introduction}

The no-short hair theorem proved in \cite{Hod11} states that the effective lengths of externally supported static 
fields (spatially regular hairy matter configurations) 
in {\it spherically} symmetric asymptotically flat black-hole spacetimes are bounded from below by the remarkably 
compact dimensionless relation
\begin{equation}\label{Eq1}
{{r_{\text{field}}}\over{r_{\text{null}}}}>1\  ,
\end{equation}
where $r_{\text{null}}$ is the radius of the innermost null circular geodesic that characterizes the hairy (non-vacuum) 
black-hole spacetime. 

Since the no-short hair lower bound (\ref{Eq1}) has rigorously been proved in the restricted regime of spherically-symmetric 
hairy black-hole spacetimes \cite{Hod11}, 
it is of physical interest to explore the regime of validity of this inequality. 
Interestingly, it has recently been proved \cite{HodKerr} that non-spherically symmetric non-static neutral 
matter configurations, which are made of stationary linearized massive scalar fields (known in the physics 
literature as scalar `clouds' \cite{Hodrc,HerR,Noteuh}) that are supported by 
spinning Kerr black holes, respect the lower bound (\ref{Eq1}). 

Intriguingly, however, it has been revealed \cite{Hodnwex} that {\it charged} scalar clouds, 
linearized charged massive scalar fields that are supported by spinning and charged Kerr-Newman black holes, 
can violate the no-short hair lower bound (\ref{Eq1}). 
In particular, it has been demonstrated \cite{Hodnwex} that stationary charged scalar clouds whose effective radii are smaller 
than the (spin and charge-dependent) radius $r_{\text{null}}(M,a,Q)$ \cite{NoteMaQ} of the black-hole null circular geodesic 
can be supported by extremal Kerr-Newman black holes.

The theorems presented in \cite{Hod11,Hodnwex,Hodnex} for the effective lengths of 
externally supported matter configurations in non-vacuum black-hole spacetimes naturally raise 
the following important question: 
How short can the composed Kerr-Newman-charged-massive-scalar-field cloudy configurations be? 
In particular, one would like to know whether a spinning and charged Kerr-Newman black hole can support 
arbitrarily compact external matter configurations of the stationary bound-state charged massive scalar fields?

The main goal of the present paper is to provide, using analytical techniques, 
explicit answers to these physically intriguing questions. 
To this end, we shall analyze below the physical and mathematical properties of the 
Klein-Gordon wave equation [see Eq. (\ref{Eq8}) below] which determines the spatio-temporal functional behaviors of charged massive scalar fields 
in the spinning and charged Kerr-Newman 
black-hole spacetime. 

Interestingly, we shall reveal the fact that, although the stationary charged scalar clouds can violate the lower bound (\ref{Eq1}) 
(which, as emphasized above, has formally been 
derived for spherically-symmetric static hairy configurations \cite{Hod11}), they {\it cannot} be arbitrarily compact. 
In particular, we shall prove that the effective proper lengths of the charged massive scalar clouds are 
characterized by the dimensionless relation 
\begin{equation}\label{Eq2}
{{\ell}\over{M}}>O(1)\  ,
\end{equation}
where $M$ is the mass of the central supporting Kerr-Newman black hole. 

Intriguingly, we shall explicitly prove that the analytically derived lower bound (\ref{Eq2}) 
[see Eq. (\ref{Eq60}) below for the exact mathematical formulation of the no-short hair lower bound] 
on the effective proper lengths of the stationary bound-state charged scalar clouds in the spinning and charged Kerr-Newman 
black-hole spacetimes is universal in the sense that it is independent of the physical parameters (dimensionless 
spin $a/M$ and dimensionless electric charge $Q/M$) of the central supporting black holes. 
Moreover, the lower bound to be derived below is generic in the sense that it is valid for all physically allowed values 
of the parameters (electric charge $q$, proper mass $\mu$, and angular harmonic indexes $\{\l,m\}$) that characterize 
the externally supported charged scalar clouds.

The structure of the paper is as follows: Section II contains a description of
the composed black-hole-scalar-field system, introducing some important quantities that will be used
later to determine the effective lengths of the supported charged clouds. 
In section III we shall explicitly prove that the radial eigenfunction that characterizes the spatial behavior of the 
charged massive scalar fields has a non-monotonic functional behavior in the Kerr-Newman black-hole spacetime. 
In particular, we shall prove that the radial scalar eigenfunction has at least one extremum point 
in the exterior region of the spinning and charged black-hole spacetime. In section IV we shall derive, using 
analytical techniques, a parameter-dependent lower bound on the effective radial lengths of
the supported charged massive scalar clouds. 

In section V we shall use the results of the previous sections in order to prove 
that the charged massive scalar clouds cannot be made arbitrarily short. 
In particular, we shall derive a generic lower bound, which is independent of the physical parameters that 
characterize the supported fields, on the effective 
radial lengths of the charged massive scalar clouds. 
In section VI we shall derive a parameter-independent lower bound on the effective proper lengths of
the charged scalar clouds in the spinning and charged Kerr-Newman black-hole spacetime. 
In section VII we shall summarize our main analytically derived results. 

Before proceeding, it is important to emphasize that bound-state massive scalar clouds extend formally to spatial 
infinity where they decay exponentially [see Eq. (\ref{Eq13}) below]. 
As a result, various criteria can be used to define their effective lengths. 
In the present study we shall derive, using {\it analytical} techniques, a generic (parameter-independent) 
lower bound on the radial peak location that characterizes the non-monotonic radial 
functional behavior of bounded charged massive scalar clouds. 
Alternatively, one can define the effective length of a scalar cloud as the radius $r_{1/2}$ that contains $50\%$ of the cloud mass. 
The determination of $r_{1/2}$ would require numerical computations which are beyond the scope of the present analytical study.  

\section{Description of the system}

We shall analyze the physical and mathematical properties of stationary bound-state linearized charged scalar clouds that 
are supported in spinning and charged Kerr-Newman black-hole spacetimes which, using the 
Boyer-Lindquist spacetime coordinates, are described by the curved line element \cite{Chan,Kerr,Newman,Noteunits,Notesmp}
\begin{eqnarray}\label{Eq3}
ds^2=-{{\Delta}\over{\rho^2}}(dt-a\sin^2\theta
d\phi)^2+{{\rho^2}\over{\Delta}}dr^2+\rho^2
d\theta^2+{{\sin^2\theta}\over{\rho^2}}\big[a
dt-(r^2+a^2)d\phi\big]^2\  .
\end{eqnarray}
The spatially-dependent metric functions in (\ref{Eq3}) are given by the functional expressions 
\begin{equation}\label{Eq4}
\Delta\equiv r^2-2Mr+a^2+Q^2\ \ \ \ ; \ \ \ \ \rho^2\equiv r^2+a^2\cos^2\theta\  ,
\end{equation}
where the physical parameters $\{M,J\equiv Ma,Q\}$ are respectively the mass, angular momentum, 
and electric charge of the black hole. 
The characteristic zeros of the metric function $\Delta$ determine the (outer and inner) 
horizon radii,
\begin{equation}\label{Eq5}
r_{\pm}=M\pm\sqrt{M^2-a^2+Q^2}\  ,
\end{equation}
of the spinning and charged black hole.

Using the mathematical decomposition
\begin{equation}\label{Eq6}
\Psi(t,r,\theta,\phi)=
\int\sum_{l,m}e^{im\phi}{S_{lm}}(\theta;m,a\sqrt{\mu^2-\omega^2})
{R_{lm}}(r;M,a,Q,\mu,q,\omega)e^{-i\omega t}d\omega\
\end{equation}
for the stationary bound-state charged massive scalar field 
(here $\omega$ is the frequency of the stationary scalar mode, $\{l,m\}$ are its 
spheroidal and azimuthal angular harmonic indexes, $\mu$ is the proper mass of the scalar field, 
and $q$ is its charge coupling constant \cite{Noteppp,Notedim}) and defining 
the dimensionless black-hole spin parameter
\begin{equation}\label{Eq7}
s\equiv {{a}\over{r_+}}\  ,
\end{equation}
one finds that the Klein-Gordon wave equation
\begin{equation}\label{Eq8}
[(\nabla^\nu-iqA^\nu)(\nabla_{\nu}-iqA_{\nu}) -\mu^2]\Psi=0\
\end{equation}
(here $A_{\nu}$ is the electromagnetic potential of the spinning and charged black hole), which determines 
the spatio-temporal functional behaviors of the charged massive scalar field in the curved black-hole spacetime, 
yields the angular differential equation \cite{Stro,Heun,Fiz1,Teuk,Abram,Hodasy,Notesa}
\begin{eqnarray}\label{Eq9}
{1\over {\sin\theta}}{{d}\over{\theta}}\Big(\sin\theta {{d S_{lm}}\over{d\theta}}\Big)
+\Big[K_{lm}+a^2(\mu^2-\omega^2)\sin^2\theta-{{m^2}\over{\sin^2\theta}}\Big]S_{lm}=0\  ,
\end{eqnarray}
which is coupled to the radial differential equation \cite{Teuk,Stro,Notert}
\begin{equation}\label{Eq10}
\Delta{{d} \over{dr}}\Big(\Delta{{dR_{lm}}\over{dr}}\Big)+\Big\{[\omega(r^2+a^2)-ma-qQr]^2
+\Delta[2ma\omega-\mu^2(r^2+a^2)-K_{lm}]\Big\}R_{lm}=0\  .
\end{equation}

The characteristic discrete set of angular eigenvalues $\{K_{lm}(a\sqrt{\mu^2-\omega^2})\}$, 
which couple the radial differential equation (\ref{Eq10}) to the angular differential equation (\ref{Eq9}), is 
determined from (\ref{Eq9}) with the physically motivated boundary condition of regularity of the angular scalar eigenfunction ${S_{lm}}(\theta;l,m,a\sqrt{\mu^2-\omega^2})$ \cite{Notebr} 
at the angular poles $\theta=0$ and $\theta=\pi$ of the black-hole 
spacetime (see \cite{Barma,Hodpp} and references therein). 
For later purposes we note that the scalar angular eigenvalues are characterized by the compact 
lower bound \cite{Barma,Notesi}
\begin{equation}\label{Eq11}
K_{lm}\geq m^2-a^2(\mu^2-\omega^2)\  .
\end{equation}

The radial equation (\ref{Eq10}) determines the spatial behavior of the stationary bound-state linearized 
charged massive scalar clouds in the spinning and charged Kerr-Newman black-hole spacetime (\ref{Eq3}). 
This differential equation is supplemented by the boundary conditions \cite{Hodrc,HerR,Ins2,Notepp}
\begin{equation}\label{Eq12}
0\leq R(r=r_+)<\infty\
\end{equation}
and [assuming $\mu^2-\omega^2>0$, see Eq. (\ref{Eq18}) below] \cite{Noteepp}
\begin{equation}\label{Eq13}
R(r\to\infty)\sim e^{-\sqrt{\mu^2-\omega^2} r}\to0\  ,
\end{equation}
which respectively correspond to a finite functional behavior of the spatially regular scalar eigenfunction on 
the outer horizon of the central supporting black hole and an exponentially decaying asymptotic radial 
behavior of the bound-state (normalizable) massive field at spatial infinity.  

The composed bound-state Kerr-Newman-black-hole-linearized-charged-massive-scalar-field configurations
\cite{Hodrc,HerR} owe their existence to the intriguing physical phenomenon of superradiant scattering 
of bosonic fields in spinning and charged black-hole spacetimes \cite{Zel,PressTeu1,Bekad}. 
In particular, for given values of the scalar field parameters $\{m,q\}$, 
the orbital frequencies of the stationary charged scalar clouds are in resonance,
\begin{equation}\label{Eq14}
\omega_{\text{field}}=\omega_{\text{c}}\  ,
\end{equation}
with the critical frequency \cite{Hodrc,Noteunits}
\begin{equation}\label{Eq15}
\omega_{\text{c}}\equiv m\Omega_{\text{H}}+q\Phi_{\text{H}}\
\end{equation}
that determines the threshold of the superradiant scattering phenomenon 
in the spinning and charged black-hole spacetime, where \cite{Chan,Kerr,Newman}
\begin{equation}\label{Eq16}
\Omega_{\text{H}}={{a}\over{r^2_++a^2}}\
\end{equation}
is the horizon angular velocity of the central Kerr-Newman black hole and
\begin{equation}\label{Eq17}
\Phi_{\text{H}}={{Qr_+}\over{r^2_++a^2}}\
\end{equation}
is its electric potential. 
Intriguingly, it has been explicitly proved \cite{Hodrc,HerR} that 
the black-hole-field resonance condition (\ref{Eq14}) allows the charged scalar clouds to coexist in a stationary 
equilibrium configuration with the central supporting Kerr-Newman black hole.

In addition to the black-hole-field resonance condition (\ref{Eq14}) that characterizes the frequency 
of the stationary scalar cloud, the proper frequency $\omega$ 
of a bound-state field (which is characterized by a normalizable wave function) of proper mass $\mu$ should be 
bounded from above by the frequency-mass relation \cite{Hodrc,HerR}
\begin{equation}\label{Eq18}
\omega^2_{\text{field}}<\mu^2\  .
\end{equation}
Taking cognizance of the asymptotic boundary condition (\ref{Eq13}), one realizes that 
the characteristic frequency-mass inequality (\ref{Eq18}) guarantees that the supported bound-state massive fields have a 
spatially bounded asymptotic functional behavior.  

\section{The effective binding potential and the near-horizon functional behavior 
of the composed black-hole-charged-massive-scalar-field system}

The radial differential equation (\ref{Eq10}) of the charged massive scalar field in the spinning and charged 
Kerr-Newman black-hole spacetime (\ref{Eq3}) can be written in the mathematically compact form 
\begin{equation}\label{Eq19}
{{d^2\psi}\over{dy^2}}-V(y)\psi=0\  ,
\end{equation}
where 
\begin{equation}\label{Eq20}
\psi=rR\ 
\end{equation}
and the radial coordinate $y$ in the Schr\"odinger-like equation (\ref{Eq19}) is 
defined by the differential relation \cite{Notemap}
\begin{equation}\label{Eq21}
dy={{r^2}\over{\Delta}}dr\  .
\end{equation}
The effective binding potential, which characterizes the composed Kerr-Newman-black-hole-charged-massive-scalar-field system, 
is given by the functional expression
\begin{equation}\label{Eq22}
V=V(r;M,a,Q,\mu,q,l,m)={{2\Delta}\over{r^6}}[Mr-(Q^2+a^2)]+{{\Delta}\over{r^4}}
[K_{lm}-2ma\omega_{\text{c}}+\mu^2(r^2+a^2)]-{{1}\over{r^4}}[\omega_{\text{c}}(r^2+a^2)-ma-qQr]^2\  .
\end{equation}

We shall now analyze the radial functional behavior of the scalar eigenfunction in the near-horizon region 
of the curved black-hole spacetime (\ref{Eq3}). 
To this end, it is convenient to define the dimensionless physical parameters \cite{NoteFF}
\begin{equation}\label{Eq23}
x\equiv {{r-r_+}\over{r_+}}\ \ \ \ ; \ \ \ \
\tau\equiv{{r_+-r_-}\over{r_+}}\ \ \ \ ; \ \ \ \ 
H\equiv K_{lm}-{{2ma(ma+qQr_+)}\over{r^2_++a^2}}+\mu^2(r^2_++a^2)>0\  ,
\end{equation}
in terms of which the Schr\"odinger-like ordinary differential equation (\ref{Eq19}) can be written, 
in the near-horizon region
\begin{equation}\label{Eq24}
x\ll\tau\  ,
\end{equation}
in the form \cite{Hodnex}
\begin{equation}\label{Eq25}
{{d^2\psi}\over{d y^2}}-{{\tau H}\over{r^2_+}}e^{{{\tau}\over{r_+}}y}\psi=0\  ,
\end{equation}
which yields the radial scalar eigenfunction \cite{Noteab1,Notesk}:
\begin{equation}\label{Eq26}
\psi(y)=I_0\Big(2\sqrt{{{H}\over{\tau}}}e^{\tau y/2r_+}\Big)\ 
\end{equation}
in the near-horizon region, 
where $I_0$ is the modified Bessel function of the first kind \cite{Abram}. 

Using the well-known mathematical properties of the modified Bessel function of the first kind
$I_0$ \cite{Abram}, one learns that the radial scalar eigenfunction $\psi(y)$ is characterized by 
the near-horizon [see Eq. (\ref{Eq24})] properties \cite{Hodnex}
\begin{equation}\label{Eq27}
\Big\{\psi>0\ \ \ ;\ \ \ {{d\psi}\over{dy}}>0\ \ \ ;\ \ \
{{d^2\psi}\over{dy^2}}>0\Big\}\ \ \ \ \text{for}\ \ \ \ 0<x\ll\tau\  .
\end{equation}

Interestingly, and most importantly for our analysis, one deduces from the near-horizon [$(r-r_+)/r_+\ll1$] 
functional behavior (\ref{Eq27}) and the large-$r$ ($r\gg r_+$) asymptotic behavior (\ref{Eq13}) of 
the radial eigenfunction $\psi$, which characterizes the spatial behavior of the 
stationary bound-state charged massive scalar fields that are supported by 
the spinning and charged Kerr-Newman black hole, that the scalar clouds have 
a non-monotonic radial behavior in the black-hole spacetime. 
In particular, the scalar eigenfunction $\psi$ possesses 
at least one extremum point, $r=r_{\text{max}}>r_+$, in the exterior region of the black-hole spacetime. 

In the next sections we shall use this important observation in order to derive, using analytical techniques, 
a universal (that is, parameter-{\it independent}) lower bound on the effective proper lengths of the stationary 
spatially regular charged scalar clouds in the spinning and charged Kerr-Newman black-hole spacetime. 

\section{Lower bound on the effective lengths of the charged scalar clouds in the Kerr-Newman black-hole spacetime}

In the previous section we have proved that the radial eigenfunction $\psi$, which determines the 
spatial behavior of the stationary bound-state charged massive scalar fields in the Kerr-Newman 
black-hole spacetime (\ref{Eq3}), must have (at least) one maximum point which is located outside the 
horizon ($r_{\text{max}}>r_+$) of the spinning and charged black hole. 
This maximum point is characterized by the functional relations
\begin{equation}\label{Eq28}
\Big\{\psi>0\ \ \ ;\ \ \ {{d\psi}\over{dy}}=0\ \ \ ;\ \ \
{{d^2\psi}\over{dy^2}}<0\Big\}\ \ \ \ \text{for}\ \ \ \ r=r_{\text{max}}\  .
\end{equation}
In the present section we shall derive a generic 
lower bound on the radial peak location $r_{\text{max}}$ that characterizes the non-monotonic radial 
functional behavior of the bound-state charged massive scalar clouds. 

To this end, we first point out that the characteristic relations (\ref{Eq27}) and (\ref{Eq28}) imply that 
the radial function $\psi$ of the charged massive scalar fields must have an
inflection point $r=r_0$ with the property 
\begin{equation}\label{Eq29}
{{d^2\psi}\over{dy^2}}=0\ \ \ \ \text{for}\ \ \ \ r=r_0\  ,
\end{equation}
which is located in the radial interval $r_0\in(r_+,r_{\text{max}})$. That is,
\begin{equation}\label{Eq30}
r_+<r_0<r_{\text{max}}\  .
\end{equation}
Taking cognizance of Eqs. (\ref{Eq19}) and (\ref{Eq29}), one realizes 
that the inflection point (\ref{Eq30}) of the scalar eigenfunction is a turning point of the 
characteristic binding potential (\ref{Eq22}) of 
the composed black-hole-charged-field system. In particular, the inflection point is determined by the simple 
functional relation
\begin{equation}\label{Eq31}
V(r=r_0)=0\  .
\end{equation}

In order to derive the lower bound on the effective proper lengths of the charged massive scalar clouds, 
we shall first derive a lower bound on the radial location of the inflection point (\ref{Eq30}). 
Taking cognizance of Eqs. (\ref{Eq11}), (\ref{Eq14}), (\ref{Eq15}), (\ref{Eq16}), (\ref{Eq17}), (\ref{Eq18}), and (\ref{Eq22}), 
one finds the (rather cumbersome) inequality \cite{Noterf,Notemn0}
\begin{eqnarray}\label{Eq32}
V(r;M,a,Q,\mu,q,l,m)>
{{m^2(r-r_+)}\over{r^4(1+s^2)^2}}\Big\{
r^2\Big[\Big({{s^2+\gamma}\over{s}}\Big)^2\cdot {{r_+-r_-}\over{r^2_+}}-{{2(1-\gamma)(s^2+\gamma)}\over{r_+}}\Big]+\nonumber \\ 
r\Big[(1-\gamma)^2(1-s^2)+2(1-\gamma)(s^2+\gamma)\Big]+
(1-\gamma)^2(s^2r_+-r_-)\Big\}\
\end{eqnarray}
for the effective binding potential of the composed Kerr-Newman-black-hole-charged-massive-scalar-field cloudy configurations, 
where we have used here the dimensionless black-hole-field physical parameter
\begin{equation}\label{Eq33}
\gamma\equiv{{qQs}\over{m}}\  .
\end{equation}

From Eqs. (\ref{Eq5}) and (\ref{Eq7}) one finds the relation
\begin{equation}\label{Eq34}
s^2r_+-r_-=-{{Q^2}\over{r_+}}\leq0\  ,
\end{equation}
which yields the inequality
\begin{equation}\label{Eq35}
s^2r_+-r_-\geq (s^2r_+-r_-)\cdot{{r}\over{r_+}}\ \ \ \ \text{for}\ \ \ \ r\geq r_+\  .
\end{equation}
Substituting (\ref{Eq35}) into (\ref{Eq32}), one obtains the inequality
\begin{eqnarray}\label{Eq36}
V(r;M,a,Q,\mu,q,l,m)>
{{m^2(r-r_+)}\over{r^3(1+s^2)^2r_+}}\Big\{
r\Big[\Big({{s^2+\gamma}\over{s}}\Big)^2\tau-2(1-\gamma)(s^2+\gamma)\Big]+\nonumber \\ 
r_+\Big[(1-\gamma)^2\tau+2(1-\gamma)(s^2+\gamma)\Big]\Big\}\  .
\end{eqnarray}
Taking cognizance of Eqs. (\ref{Eq31}) and (\ref{Eq36}), one finds the characteristic inequality
\begin{eqnarray}\label{Eq37}
{{m^2(r_0-r_+)}\over{r^3_0(1+s^2)^2r_+}}\Big\{
r_0\Big[\Big({{s^2+\gamma}\over{s}}\Big)^2\tau-2(1-\gamma)(s^2+\gamma)\Big]+
r_+\Big[(1-\gamma)^2\tau+2(1-\gamma)(s^2+\gamma)\Big]\Big\}<0\ 
\end{eqnarray}
for the innermost inflection point $r=r_0$ of the radial scalar eigenfunction.

For later purposes, it is important to point out that 
\begin{equation}\label{Eq38}
(1-\gamma)(s^2+\gamma)>0\  .
\end{equation}
In order to prove the inequality (\ref{Eq38}), one may rewrite Eq. (\ref{Eq37}) in the form
\begin{eqnarray}\label{Eq39}
{{m^2(r_0-r_+)}\over{r^3_0(1+s^2)^2r_+}}\Big\{
\tau\Big[r_0\Big({{s^2+\gamma}\over{s}}\Big)^2+r_+(1-\gamma)^2\Big]+
2(1-\gamma)(s^2+\gamma)(r_+-r_0)\Big\}<0\  .
\end{eqnarray}
Substituting into (\ref{Eq39}) the characteristic inequalities 
\begin{equation}\label{Eq40}
r_0\Big({{s^2+\gamma}\over{s}}\Big)^2+r_+(1-\gamma)^2>0\ \ \ \ ;\ \ \ \ r_+-r_0\leq0\  ,
\end{equation}
one obtains the dimensionless inequality (\ref{Eq38}). 
In addition, substituting the inequality [see Eq. (\ref{Eq38})] 
\begin{eqnarray}\label{Eq41}
(1-\gamma)^2\tau+2(1-\gamma)(s^2+\gamma)\geq2(1-\gamma)(s^2+\gamma)>0
\end{eqnarray}
into Eq. (\ref{Eq37}), one finds the characteristic relation 
\begin{eqnarray}\label{Eq42}
\Big({{s^2+\gamma}\over{s}}\Big)^2\tau-2(1-\gamma)(s^2+\gamma)<0\  .
\end{eqnarray}

Taking cognizance of Eqs. (\ref{Eq30}), (\ref{Eq37}), and (\ref{Eq42}), 
one obtains the dimensionless lower bound 
\begin{equation}\label{Eq43}
{{r_{\text{max}}}\over{r_+}}>{{r_0}\over{r_+}}>{{2(1-\gamma)(s^2+\gamma)+(1-\gamma)^2\tau}\over
{2(1-\gamma)(s^2+\gamma)-\big({{s^2+\gamma}\over{s}}\big)^2\tau}}\
\end{equation}
on the radial location of the extremum point which characterizes the eigenfunction $\psi$ of 
the stationary bound-state charged scalar fields. 
The analytically derived inequality (\ref{Eq43}) provides a lower bound on the effective radial lengths of the charged massive 
scalar clouds in the supporting Kerr-Newman black-hole spacetime. 

\section{Generic (parameter-independent) bound on the effective lengths of the charged scalar clouds}

Our main goal is to derive, using analytical techniques, a universal lower bound on the effective lengths of 
the stationary bound-state charged scalar clouds that are supported by 
the spinning and charged Kerr-Newman black-hole spacetime (that is, 
a bound which is {\it independent} of the physical parameters $\{q,\mu,l,m\}$ 
that characterize the externally supported spatially regular charged massive scalar fields). 
To this end, it proves useful to define the dimensionless variable \cite{Notebp}
\begin{equation}\label{Eq44}
\beta\equiv {{1-\gamma}\over{s^2+\gamma}}>0\  , 
\end{equation}
in terms of which the inequality (\ref{Eq43}) can be written in the form 
\begin{equation}\label{Eq45}
{{r_0}\over{r_+}}>F(\beta)\  ,
\end{equation}
where \cite{Notebpp}
\begin{equation}\label{Eq46}
F(\beta)\equiv {{1+{1\over2}\beta\cdot\tau}\over{1-{{1}\over{2s^2\beta}}\cdot\tau}}\  .
\end{equation}

In order to derive the generic (that is, independent of the physical parameters $\{q,\mu,l,m\}$ of the scalar field) 
bound, we shall now determine the particular value of the composed dimensionless 
parameter $\beta=\beta_{\text{max}}$ that, for given values $\{\tau,s\}$ of the dimensionless physical parameters that 
characterize the central supporting Kerr-Newman black hole, minimizes the value of the function $F(\beta)$ 
on the right-hand-side of the dimensionless lower bound (\ref{Eq45}). 
Substituting (\ref{Eq46}) into the characteristic relation
\begin{equation}\label{Eq47}
{{dF(\beta)}\over{d\beta}}=0\ \ \ \ \text{for}\ \ \ \ \beta=\beta_{\text{max}}\  ,
\end{equation}
one finds the functional expression \cite{Notebetm}
\begin{equation}\label{Eq48}
\beta_{\text{max}}={{\tau}\over{2s^2}}\Big[1+\sqrt{1+\Big({{2s}\over{\tau}}\Big)^2}\Big]\  .
\end{equation}
Substituting (\ref{Eq48}) into (\ref{Eq46}) and using the inequality (\ref{Eq45}), one obtains the lower bound
\begin{equation}\label{Eq49}
{{r_0}\over{r_+}}>1+{{2\big(1+\sqrt{1+\eta}\big)}\over{\eta}}\
\end{equation} 
on the radial location of the inflection point (\ref{Eq30}) that characterizes the scalar eigenfunction, where 
\begin{equation}\label{Eq50}
\eta\equiv \Big({{2s}\over{\tau}}\Big)^2\  .
\end{equation}

Taking cognizance of the characteristic radial inequalities (\ref{Eq30}) and
(\ref{Eq49}), one finds the remarkably compact dimensionless lower bound [see Eq. (\ref{Eq23})]
\begin{equation}\label{Eq51}
x_{\text{max}}>{{2\big(1+\sqrt{1+\eta}\big)}\over{\eta}}\  
\end{equation}
on the radial location of the extremum point which characterizes the radial 
eigenfunction $\psi$ of the supported spatially regular stationary charged massive scalar fields in the spinning and 
charged Kerr-Newman black-hole spacetime. 

It is physically interesting to stress the fact that the analytically derived inequality (\ref{Eq51}) 
provides a parameter-independent bound on the effective lengths of the supported charged massive scalar clouds. 
In particular, the lower bound (\ref{Eq51}) is universal in the sense that it is {\it independent} 
of the physical parameters (electric charge $q$, proper mass $\mu$, and angular
harmonic indexes $\{l,m\}$) that characterize the spatially regular bound-state charged massive scalar fields.

\section{Generic lower bound on the effective proper lengths of the stationary bound-state charged scalar clouds}

In the present section we shall explicitly prove that the dimensionless {\it proper} lengths $\ell/M$ 
of the stationary bound-state charged massive scalar clouds 
can be bounded from below by a universal relation which does not depend on any of the 
physical parameters $\{M,a,Q,q,\mu,l,m\}$ that characterize the 
composed Kerr-Newman-black-hole-charged-massive-scalar-field system. 

To this end, we first point out that the right-hand-side of the analytically derived lower bound (\ref{Eq51}), 
which characterizes the spatial behavior of the composed Kerr-Newman-charged-massive-scalar-field bound-state 
configurations, is a monotonically decreasing function of the dimensionless variable $\eta$ with 
the asymptotic property $[2\big(1+\sqrt{1+\eta}\big)]/\eta\to 2/\sqrt{\eta}=\tau/s$ [see Eq. (\ref{Eq50})] 
for $\eta\to\infty$. This fact implies that the 
effective lengths of the charged clouds are bounded from below by the inequality  
\begin{equation}\label{Eq52}
x_{\text{max}}>{{\tau}\over{s}}\  .  
\end{equation}

The proper distance of the maximum point (\ref{Eq52}) to the black-hole horizon is given by the 
integral relation \cite{Bekpd}
\begin{equation}\label{Eq53}
\ell_{\text{max}}=\int_{r_+}^{r_{\text{max}}}\sqrt{g_{rr}}dr\  ,
\end{equation}
which, using (\ref{Eq3}), yields the inequality \cite{Noteeq}
\begin{equation}\label{Eq54}
\ell_{\text{max}}=\int_{r_+}^{r_{\text{max}}}\sqrt{{{r^2+a^2\cos^2\theta}\over{r^2-2Mr+a^2+Q^2}}}\ dr\geq
\int_{r_+}^{r_{\text{max}}}\sqrt{{{1}\over{1-{{2M}\over{r}}+{{a^2+Q^2}\over{r^2}}}}}\ dr\  .
\end{equation}
Interestingly, and most importantly for our analysis, 
the integral on the right-hand-side of (\ref{Eq54}) can be evaluated analytically to yield the functional expression 
\begin{equation}\label{Eq55}
\ell_{\text{max}}\geq \sqrt{r^2_{\text{max}}-2Mr_{\text{max}}+a^2+Q^2}+
M\cdot \cosh^{-1}\Big({{r_{\text{max}}-M}\over{r_+-M}}\Big)\  .
\end{equation}
Taking cognizance of Eq. (\ref{Eq23}), one can write the inequality (\ref{Eq55}) in the form
\begin{equation}\label{Eq56}
\ell_{\text{max}}\geq r_+\cdot \sqrt{x_{\text{max}}(x_{\text{max}}+\tau)}+
M\cdot \cosh^{-1}\Big({{2x_{\text{max}}+\tau}\over{\tau}}\Big)\  .
\end{equation}
Substituting the inequality (\ref{Eq52}) into (\ref{Eq56}), one obtains the remarkably compact relation \cite{Notemnn}
\begin{equation}\label{Eq57}
\ell_{\text{max}}> r_+\tau\cdot \sqrt{{{1+s}\over{s^2}}}+
M\cdot \cosh^{-1}\Big(1+{{2}\over{s}}\Big)\  .
\end{equation}

Intriguingly, one finds that the right-hand-side of the inequality (\ref{Eq57}) cannot be made arbitrarily small. 
In particular, it can be minimized by the limiting values $\tau\to0^+$ with $s\to1^-$ 
to yield the universal (parameter-{\it independent}) lower bound \cite{Noteln}
\begin{equation}\label{Eq58}
\ell_{\text{max}}>M\cdot\ln(3+\sqrt{8})\
\end{equation}
on the effective proper lengths of the stationary bound-state charged scalar clouds in the spinning and charged 
Kerr-Newman black-hole spacetime. 

\section{Summary}

Recent analytical \cite{Hodrc} and numerical \cite{HerR} studies have revealed the physically interesting fact that 
the influential no-hair conjecture can be violated by hairy black-hole solutions of the composed 
Einstein-Maxwell-minimally-coupled-charged-massive-scalar field theory. 
In particular, it has been explicitly proved that asymptotically flat spinning black holes can support spatially 
regular matter configurations which are made of stationary (neutral or charged) massive scalar fields \cite{Hodrc,HerR}. 

Intriguingly, it has been proved \cite{Hodnwex} that spinning and charged Kerr-Newman black-hole solutions of the 
Einstein-Maxwell-scalar field theory 
can support non-spherically symmetric stationary bound-state charged scalar clouds whose effective lengths violate the 
no-short hair lower bound (\ref{Eq1}) which has rigorously been proved for 
spherically-symmetric hairy black-hole configurations \cite{Hod11}.

Motivated by the analytical results presented in \cite{Hod11,Hodnwex,Hodnex} for the effective lengths of field 
configurations that are supported in black-hole spacetimes, we have raised 
the following physically interesting question: 
How short can the composed Kerr-Newman-charged-massive-scalar-field cloudy configurations be? 

In order to address this intriguing physical question, we have analyzed the radial functional behavior 
of the (rather cumbersome) effective binding potential [see Eq. (\ref{Eq22})] that 
characterizes the composed Kerr-Newman-black-hole-linearized-charged-massive-scalar-field 
cloudy configurations. 
The main {\it analytical} results derived in this paper and their physical implications are as follows:

(1) We have explicitly proved that, although stationary charged scalar clouds in the spinning and charged 
Kerr-Newman black-hole spacetime can violate the no-short hair lower bound (\ref{Eq1}) (which is formally 
valid for spherically symmetric hairy black-hole spacetimes), they cannot be made arbitrarily short. 

(2) In particular, it has been proved that the scalar eigenfunction $\psi$, which characterizes the spatial behavior of the 
stationary bound-state charged massive scalar clouds in the spinning and charged Kerr-Newman black-hole spacetime (\ref{Eq3}), 
has a non-trivial (non-monotonic) radial functional behavior [see Eq. (\ref{Eq28})]. 

(3) We have proved that the effective radial lengths of the bound-state charged massive scalar clouds are 
bounded from below by the functional relation [see Eqs. (\ref{Eq23}), (\ref{Eq50}), and (\ref{Eq51})]
\begin{equation}\label{Eq59}
{{r_{\text{max}}-r_+}\over{r_+}}>{{2\Big[1+\sqrt{1+\big({{2s}
\over{\tau}}\big)^2}\Big]}\over{\big({{2s}\over{\tau}}\big)^2}}\  ,
\end{equation}
where $r_{\text{max}}$ is the radial location of the maximum point which characterizes the spatially 
regular non-monotonic scalar eigenfunction $\psi$ of the charged clouds. 

It is worth emphasizing again that the bound-state massive scalar clouds extend formally to spatial 
infinity [where they are characterized by exponentially decaying radial eigenfunctions, see Eq. (\ref{Eq13})] and one can 
therefore use various criteria to define their effective lengths. 
For example, one can define the effective length of a scalar cloud as the radius $r_{1/2}$ that contains $50\%$ of its mass. 
Determining $r_{1/2}$ requires numerical computations which are beyond the scope of the present analytical study.  

It is worth emphasizing the fact that the analytically derived lower bound (\ref{Eq59}) on the 
effective lengths of the charged scalar clouds is of general validity in the sense 
that it does not depend on the physical parameters $\{q,\mu,l,m\}$ of the externally supported fields.
  
(4) Using the analytically derived functional relation (\ref{Eq59}), we have 
derived the dimensionless lower bound 
\begin{equation}\label{Eq60}
{{\ell_{\text{max}}}\over{M}}>\ln(3+\sqrt{8})\
\end{equation}
on the effective proper lengths of the stationary bound-state charged scalar clouds in the Kerr-Newman 
black-hole spacetime. 

Intriguingly, the analytically derived bound (\ref{Eq60}) \cite{Notefn} on the proper lengths of the charged clouds 
is {\it universal} in the sense that it is valid 
for all Kerr-Newman black-hole spacetimes and for all values of the physical parameters 
(electric charge $q$, proper mass $\mu$, and angular harmonic indexes $\{\l,m\}$) that characterize 
the stationary bound-state charged massive scalar fields.

\bigskip
\noindent
{\bf ACKNOWLEDGMENTS}
\bigskip

This research is supported by the Carmel Science Foundation. I thank
Yael Oren, Arbel M. Ongo, Ayelet B. Lata, and Alona B. Tea for
stimulating discussions.



\begin{thebibliography}{99}
\bibitem{Noteuh} Note that the term `scalar clouds' is usually used in the physics literature to
describe scalar fields that are linearly coupled to the supporting black hole \cite{Hodrc,HerR}, 
whereas the term `scalar hair' is usually used in the physics literature to
describe scalar fields with a non-linear coupling to the supporting black-hole spacetime.
\bibitem{NoteMaQ} The physical parameters $\{M,a,Q\}$ are respectively the mass, angular momentum per unit mass, 
and electric charge of the supporting Kerr-Newman black hole.
\bibitem{Noteunits} We shall use natural units in which $G=c=\hbar=1$.
\bibitem{Notesmp} We shall henceforth assume, without loss of generality, the relations $a>0$ and $Q\geq0$ for the 
physical parameters of the supporting Kerr-Newman black hole.
\bibitem{Noteppp} We shall henceforth assume 
the relations $m>0$ and 
$q\geq0$ for the physical parameters of the supported charged massive scalar field.
\bibitem{Notedim} Note that the physical parameters $\{\mu,q\}$ that characterize the charged massive scalar field 
stand respectively for $\{\mu/\hbar,q/\hbar\}$ and therefore (see \cite{Noteunits}) have the dimensions of $($length$)^{-1}$.
\bibitem{Notesa} Note that the angular functions ${S_{lm}}(\theta;m,a\sqrt{\mu^2-\omega^2})$ are usually called 
the spheroidal harmonic eigenfunctions in the physics literature.
\bibitem{Notert} Note that the radial equation (\ref{Eq10}) is also known in the physics literature 
as the radial Teukolsky equation \cite{Teuk,Stro}.
\bibitem{Notebr} For brevity, we shall henceforth omit the dimensionless angular indexes $\{l,m\}$ from 
the radial and angular scalar eigenfunctions.
\bibitem{Notesi} It is interesting to point out 
that the characteristic lower bound (\ref{Eq11}) on the angular scalar eigenvalues of the ordinary 
differential equation (\ref{Eq9}) can be approached in the asymptotic $l=m\gg1$ regime \cite{Hodpp}. 
In particular, one finds the asymptotic functional behavior 
$K_{mm}(a\sqrt{\mu^2-\omega^2})=m^2[1+O(m^{-1})]-a^2(\mu^2-\omega^2)$ \cite{Hodpp}.
\bibitem{Notepp} Note that one may choose $R(r=r_+)\geq0$ for the radial scalar eigenfunction without loss of generality.
\bibitem{Noteepp} Note that one may choose $\sqrt{\mu^2-\omega^2}>0$ in the bound-state 
regime $\omega^2<\mu^2$ [see Eq. (\ref{Eq18})] without loss of generality.
\bibitem{Notemap} Note that the differential relation (\ref{Eq21}) maps the Boyer-Lindquist 
radial coordinate $r\in [r_+,\infty]$ into the regime $y\in [-\infty,+\infty]$.
\bibitem{Noteab1} We have used here equations (9.1.54) and (9.6.3) of \cite{Abram}. 
In addition, we have substituted, without loss of generality, the value $1$ for 
the normalization constant of the linearized scalar eigenfunction (\ref{Eq26}).
\bibitem{Notesk} Note that the mathematical solution (\ref{Eq26}) of the near-horizon radial 
differential equation (\ref{Eq25}) respects the physically motivated horizon 
boundary condition (\ref{Eq12}). 
The second mathematical solution of the Schr\"odinger-like ordinary differential equation (\ref{Eq25}) 
is given by 
$\psi_2(y)=K_0(2\sqrt{{{H}/{\tau}}}e^{\tau y/2r_+})$, where $K_0$ is the modified
Bessel function of the second kind. However, this mathematical solution does not
respect the boundary condition (\ref{Eq12}) [see Eq. (9.6.13) of \cite{Abram}].
\bibitem{Noterf} Here we have used the characteristic inequality $Mr-(a^2+Q^2)\geq0$ for $r\geq r_+$ [see Eq. (\ref{Eq5})].
\bibitem{Notebp} Note that the inequality $\beta>0$ follows from Eq. (\ref{Eq38}).
\bibitem{Notebetm} Note that the second solution of Eq. (\ref{Eq47}), 
$\beta={{\tau}\over{2s^2}}\Big[1-\sqrt{1+\Big({{2s}\over{\tau}}\Big)^2}\Big]<0$, violates the characteristic 
inequality (\ref{Eq44}).
\bibitem{Noteeq} Note that the equality sign in (\ref{Eq54}) corresponds to the 
proper distance in the equatorial plane of the black hole.
\bibitem{Notemnn} Here we have used the fact that $\cosh^{-1}(z)$ is a monotonically increasing function 
of its argument \cite{Abram}.
\bibitem{Noteln} Here we have used the relation $\cosh^{-1}(3)=\ln(3+\sqrt{8})$.
\bibitem{Notefn} It would be physically interesting to use sophisticated numerical 
tools \cite{HerR} in order to test the validity of our analytically derived results in the non-linear regime (in which 
case the supported charged massive scalar fields are non-linearly coupled to the curved black-hole spacetime).
\end{thebibliography}
\end{document}